\documentclass{JHEP3}

\usepackage{amsmath}
\usepackage{cite}

\def\be{\begin{equation}}
\def\ee{\end{equation}}
\def\bea{\begin{eqnarray}}
\def\eea{\end{eqnarray}}

\title{On the origin of neutrino flavour symmetry}

\author{Stephen F. King\\
School of Physics and Astronomy, University of Southampton,\\
Southampton, SO17 1BJ, U.K.\\
E-mail: \email{king@soton.ac.uk}}

\author{Christoph Luhn\\
School of Physics and Astronomy, University of Southampton,\\
Southampton, SO17 1BJ, U.K.\\
E-mail: \email{christoph.luhn@soton.ac.uk}}

\preprint{arXiv:0908.1897}

\abstract{We study classes of models which are based on some discrete family
symmetry which is completely broken such that the observed
neutrino flavour symmetry emerges indirectly as an accidental
symmetry. For such ``indirect'' models we discuss the D-term
flavon vacuum alignments which are required for such an accidental
flavour symmetry consistent with tri-bimaximal lepton mixing to
emerge. We identify large classes of suitable discrete family symmetries,
namely the $\Delta(3n^2)$ and $\Delta(6n^2)$ groups, together with
other examples such as $Z_7\rtimes Z_3$.
In such indirect models the implementation of the type I see-saw
mechanism is straightforward using constrained sequential
dominance. However the accidental neutrino flavour symmetry may be
easily violated, for example leading to a large reactor angle,
while maintaining accurately the tri-bimaximal solar and
atmospheric predictions.}


\begin{document}

\section{Introduction}
It is well known that the solar and atmospheric neutrino data are
consistent with so-called tri-bimaximal (TB) mixing \cite{HPS},
\begin{eqnarray}
U_{TB} = \left( \begin{array}{rrr}
-\frac{2}{\sqrt{6}}   & \frac{1}{\sqrt{3}} & 0 \\
\frac{1}{\sqrt{6}}  & \frac{1}{\sqrt{3}} & \frac{1}{\sqrt{2}} \\
\frac{1}{\sqrt{6}}  & \frac{1}{\sqrt{3}} & -\frac{1}{\sqrt{2}}
\end{array}
\right). \label{MNS0}
\end{eqnarray}
The ansatz of TB lepton mixing matrix is interesting due to its
symmetry properties which seem to call for a possibly discrete
non-Abelian family symmetry in nature \cite{Harrison:2003aw}.
There has been a considerable amount of theoretical work in this
direction \cite{Ma:2007wu,King:2005bj,deMedeirosVarzielas:2005ax,King:2006me,deMedeirosVarzielas:2005qg,deMedeirosVarzielas:2006fc,King:2006np,Altarelli:2006kg,Chen:2009um,Frampton:2004ud,Luhn:2007sy}.
In the neutrino flavour basis (i.e. diagonal charged lepton mass basis),
it has been shown that the TB neutrino mass matrix is invariant under
$S,U$ transformations \cite{Lam:2008sh} \be
{M^{\nu}_{TB}}\,= S {M^{\nu}_{TB}} S^T\,=
U {M^{\nu}_{TB}} U^T \ . \label{S} \ee
A very straightforward argument \cite{King:2009mk} shows
that this neutrino flavour symmetry group has only four elements
corresponding to Klein's four-group $Z_2^S \times Z_2^U$.
By contrast the diagonal charged lepton mass
matrix (in this basis) satisfies a diagonal phase
symmetry $T$. The matrices $S,T,U$ form the generators of the
group $S_4$ in the triplet representation, while the $A_4$
subgroup is generated by $S,T$.

Recently there have been two apparently conflicting claims in the
literature concerning the nature of the underlying family symmetry
$G_f$ of the Lagrangian responsible for the TB lepton mixing
matrix. It was originally claimed that $S_4$ is the minimal family
symmetry describing leptons with TB mixing \cite{Lam:2008sh}.
However this claim has recently been challenged in
\cite{Grimus:2009pg} where it is argued that TB lepton mixing
could arise from many possible candidate family symmetries which
need not contain $S_4$ as a subgroup. Most recently the original
claim has been clarified to include a discussion of $A_4$ and
$S_3$ in addition to $S_4$ \cite{Lam:2009hn}.

In this paper, motivated by the above debate,
we discuss the relation between the observed flavour
symmetry of the neutrino mass matrix $Z_2^S \times Z_2^U$ and the
underlying family symmetry of the Lagrangian $G_f$. We show that
the flavour symmetry of the neutrino mass matrix may originate
from two quite distinct classes of models. The first class of models,
which we call direct models, are based on an $A_4$ or $S_4$ family
symmetry, some of whose generators are directly preserved in the
lepton sector and are manifested as part of the observed flavour
symmetry. The second class of models, which we call indirect
models, are based on any family symmetry $G_f$ which is completely
broken in the neutrino sector, while the observed neutrino flavour
symmetry $Z_2^S \times Z_2^U$ in the neutrino flavour basis emerges as an
accidental symmetry which
is an indirect effect of the family symmetry $G_f$. In such
indirect models the flavons responsible for the neutrino masses break $G_f$
completely so that none of the generators of $G_f$ survive in the
observed flavour symmetry $Z_2^S \times Z_2^U$.

In the direct models, the symmetry of the neutrino mass matrix in
the neutrino flavour basis (henceforth called the neutrino mass matrix for
brevity) is a remnant of the $S_4$ symmetry of the Lagrangian. For
direct models, it is correct to say \cite{Lam:2008sh} that the
Lagrangian should contain $S_4$ as a subgroup, where the generators
$S,U$ are preserved in the neutrino sector, while the diagonal generator
$T$ is preserved in the charged lepton sector.
For example $PSL(2,7) = \Sigma (168)$ \cite{King:2009mk} contains
$S_4$ as a subgroup.
If the family symmetry of the underlying
Lagrangian is $A_4$, then in some cases this can lead to a direct
model where the $T$ generator of the underlying Lagrangian
symmetry is preserved in the charged lepton sector, while the $S$
generator is preserved in the neutrino sector, with the $U$
transformation of $S_4$ emerging as an accidental symmetry   due to the
absence of flavons in the ${\bf 1',1''}$ representations of $A_4$
\cite{Altarelli:2006kg}. Typically direct models satisfy form
dominance \cite{Chen:2009um}, and require flavon F-term
vacuum alignment, permitting an $SU(5)$ type unification
\cite{Altarelli:2006kg}.

The main focus of this paper is on the indirect models in which
the underlying family symmetry of the Lagrangian $G_f$ is
completely broken, and the flavour symmetry of the neutrino mass
matrix $Z_2^S \times Z_2^U$ emerges entirely as an accidental
symmetry, due to the presence of flavons with particular vacuum
alignments proportional to the columns of $U_{TB}$, where such
flavons only appear quadratically in effective Majorana
Lagrangian. We emphasise that such vacuum alignments can be
elegantly achieved using D-term vacuum alignment, and catalogue
the possible choices of discrete family symmetry $G_f$ which are
consistent with this mechanism,  namely the $\Delta(3n^2)$ and
$\Delta(6n^2)$ groups, together with other examples such as
$Z_7\rtimes Z_3$. Although the presence of the underlying family
symmetry $G_f$ is crucial for producing such vacuum alignments, we
shall show that the neutrino flavour symmetry $Z_2^S\times
Z_2^U$ in the neutrino flavour basis does not arise as a subgroup of
$G_f$ but rather accidentally.
However, since the family symmetry $G_f$ only partly enforces
the required vacuum alignments, the
$Z_2^S \times Z_2^U$ flavour symmetry can be easily
violated, possibly leading to a large reactor angle while accurately
preserving the tri-bimaximal solar and atmospheric predictions.
The see-saw mechanism can be implemented in indirect models using
Constrained Sequential Dominance (CSD)
\cite{King:2005bj,King:1998jw} and such models typically permit $SO(10)$ type unification
\cite{King:2006me}.

In section 2 we give a novel derivation of the flavour symmetry
of the neutrino mass matrix which enables the flavons of the
indirect models to be readily identified.
In section 3 we briefly discuss flavons of the direct models, then
introduce the flavons of the indirect models. In section 4 we
show how D-term vacuum alignment may be used for indirect models,
and identify the possible classes of family
symmetry which are consistent with a particularly simple and useful term in the flavon potential.
Section 5 shows how the type I see-saw mechanism can be applied
to indirect models using CSD. Section 6 concludes
the paper.


\section{The flavour symmetry of the neutrino mass matrix}
In this section we give a novel derivation of the flavour symmetry
of the neutrino mass matrix in the neutrino flavour basis which enables
the flavons of indirect models to be easily identified.
In the neutrino flavour basis, in which the charged lepton mass
matrix is diagonal and the TB mixing arises from the neutrino
sector, the effective neutrino mass matrix, denoted by
${M^{\nu}_{TB}}$, may be diagonalised as,
\begin{equation}
{M^{\nu}_{\mbox{\scriptsize diag}}} = U_{TB}^{T}
{M^{\nu}_{TB}} U_{TB}=\mathrm{diag}\,(m_{1}, \; m_{2}, \;
m_{3}) \; .
\end{equation}
Given $U_{\mbox{\scriptsize TB}}$, this enables
${M^{\nu}_{TB}}$ to be determined in terms of neutrino
masses,
\begin{equation}\label{eq:csd-tbm0}
{M^{\nu}_{TB}}= m_{1} \Phi_{1}\Phi_{1}^{T} + m_{2}
\Phi_{2}\Phi_{2}^{T} + m_{3} \Phi_{3}\Phi_{3}^{T} \; ,
\end{equation}
corresponding to the orthonormal column vectors
\begin{equation}
\label{Phi0} {\Phi}_{1}=\frac{1}{\sqrt{6}} \left(
\begin{array}{c}
-2 \\
1 \\
1
\end{array}
\right), \ \ {\Phi}_{2}=\frac{1}{\sqrt{3}} \left(
\begin{array}{c}
1 \\
1 \\
1
\end{array}
\right), \ \ {\Phi}_{3}=\frac{1}{\sqrt{2}} \left(
\begin{array}{c}
0 \\
1 \\
-1
\end{array}
\right),
\end{equation}
which are just equal to the columns of $U_{TB}$, \be
U_{TB}=(\Phi_1, \Phi_2 , \Phi_3) \label{columns} \ee with the
orthonormality relations, \be \Phi_i^T \Phi_j = \delta_{ij}.
\label{orthonormal} \ee It is convenient to define the matrices
\be G_i= \Phi_i \Phi_i^T, \ee in terms of which the neutrino mass
matrix is simply written, \be {M^{\nu}_{TB}}= m_1G_1 +
m_2G_2 +  m_3G_3, \ee and \bea
U_{TB}^T G_1 U_{TB} &=& \mathrm{diag} \left( 1,0,0 \right),\nonumber \\
U_{TB}^T G_2 U_{TB} &=& \mathrm{diag} \left( 0,1,0 \right), \\
U_{TB}^T G_3 U_{TB} &=& \mathrm{diag} \left( 0,0,1 \right). \nonumber \eea

We aim to find the symmetry transformations $G$ which leave the
TB neutrino mass matrix in Eq.\,(\ref{eq:csd-tbm0}) invariant, as in
Eq.\,(\ref{S}), which implies that, \be G \Phi_i \Phi_i^T G^T=\Phi_i
\Phi_i^T.  \label{S2} \ee Eq.\,(\ref{S2}) implies that \be G\Phi_i =
\eta_i \Phi_i, \label{G} \ee where $\eta_i= \pm 1$. For a given
choice of $\eta_i$ we can easily find the corresponding real
orthogonal matrix $G$, since it is diagonalised by the matrix
$U_{TB}= \left(\Phi_1,\Phi_2,\Phi_3\right)$ whose columns are the
eigenvectors $\Phi_i$ with eigenvalues $\eta_i$, \be
U_{TB}^TGU_{TB}={\rm diag}\left(\eta_1,\eta_2,\eta_3\right).
\label{G3} \ee Then, by inverting Eq.\,(\ref{G3}), we find a set of
eight choices of $G$ corresponding to the set of choices of
$\eta_i$, \be G \in  G_{\eta_1, \eta_2, \eta_3}\equiv
\eta_1G_1+\eta_2G_2+\eta_3G_3. \label{G2} \ee For example,
$G_{+++}=I$, the unit matrix, and we can define $S=G_{-+-}$,
$U=G_{--+}$, which are explicitly given as,
\begin{equation}
S = \frac{1}{3} \left(\begin{array}{ccc}
-1& 2  & 2  \\
2  & -1  & 2 \\
2 & 2 & -1
\end{array}\right), \qquad
U = - \left( \begin{array}{ccc}
1 & 0 & 0 \\
0 & 0 & 1 \\
0 & 1 & 0
\end{array}\right) \ .\label{SUgens}
\end{equation}
We can write \be S=-I+2G_2, \ \ U= -I+2G_3. \ee In general the
multiplication law for the $G_{\eta_1, \eta_2, \eta_3}$ is
extremely simple, \be G_{\eta_1, \eta_2, \eta_3}G_{\eta'_1,
\eta'_2, \eta'_3}=G_{\eta_1\eta'_1, \eta_2\eta'_2, \eta_3\eta'_3}
\label{product}. \ee Eq.\,(\ref{product}) follows from \be
G_iG_j=\delta_{ij}G_i, \ee which follows from the orthonormality
relations in Eq.\,(\ref{orthonormal}), as does, \be G_i\Phi_j=
\delta_{ij}\Phi_i. \label{GPhi} \ee From Eq.\,(\ref{product}) it is
easy to see that $S,U$ form a four element group together with the
unit matrix $I$, with the fourth element $X=G_{+--}$ being given
by $X=SU=US$, where all elements have determinant given by
$\eta_1\eta_2\eta_3$ equal to plus one. The remaining four
transformations with negative determinant given by $-I,-S,-U,-X$
clearly do not form a group. To summarise, Klein's four-group
$Z_2^S \times Z_2^U$ is the flavour symmetry of the TB neutrino
mass matrix with elements \be G\in (
G_{+++},G_{-+-},G_{--+},G_{+--} )\equiv ( I,S,U,X ) \label{GG} \ee
in the notation of Eq.\,(\ref{G2}).


\section{Flavons of direct and indirect models}
\subsection{The flavour symmetry problem}
The typical Lagrangian (or superpotential) of interest generically
consists of two parts, the Yukawa sector and the Majorana sector.
The Yukawa sector is of the form, \be {\mathcal L}^{Yuk} \sim
\psi_{i}Y^{Yuk}_{ij}\psi_{j}^{c}H \ , \label{opYuk} \ee while the
effective Majorana sector is of the form \be {\mathcal
L}^{Maj} \sim \psi_{i} Y^{Maj}_{ij}\psi_{j}HH \ ,
\label{opMaj} \ee where $Y^{Yuk}_{ij}$ and $Y^{Maj}_{ij}$ are
Yukawa and Majorana couplings, respectively, while $H$ are Higgs
fields. When the Higgs develop their VEVs, and $\psi$ are
identified with left-handed lepton fields $L$, while $\psi^{c}$
are identified with charged conjugated right-handed charged
leptons such as $e^c$, the Yukawa operators lead to the charged
lepton mass matrix $M^e_{ij}\propto Y^e_{ij}$, while the effective
Majorana operators lead to a neutrino Majorana mass matrix
$M^{\nu}_{ij}\propto Y^{\nu}_{ij}$.

We have already seen that the TB neutrino mass matrix is invariant under
$S,U$ transformations in Eq.\,(\ref{SUgens})
while the diagonal charged lepton mass matrix is invariant under
the phase transformation $T={\rm diag}(1,\omega^2,\omega)$ where
$\omega =e^{2\pi i/3}$. At first sight it appears paradoxical that
${\mathcal L}^{Yuk}$ and ${\mathcal L}^{Maj}$ could respect different flavour
symmetries since the lepton doublets $L$ are common to both.
Indeed, with $Y^{Yuk}_{ij}$ and $Y^{Maj}_{ij}$ being simply numbers,
a symmetry transformation $L \rightarrow V L$ in one sector would entail an
identical symmetry transformation in the other. Thus $V$ would be a symmetry
of both the neutrino and the charged lepton mass matrix.

The resolution to this problem is intrinsically related to the
origin of the Yukawa couplings. If the Yukawa couplings are
generated dynamically by the VEVs of flavon fields,
then it is possible to have different symmetries
in the Yukawa and Majorana sectors.
The idea is that the
complete high energy theory Lagrangian, including both ${\mathcal
L}^{Yuk}$ and ${\mathcal L}^{Maj}$, would both respect some
family symmetry $G_{f}$ due to the presence of flavons
$\phi^{Yuk}$ and $\phi^{Maj}$, where $\phi^{Yuk}$  appears in
${\mathcal L}^{Yuk}$ while $\phi^{Maj}$ appears in ${\mathcal
L}^{Maj}$, \bea {\mathcal L}^{Yuk} &\sim&
\psi^{}_{}\phi_{}^{Yuk}\psi_{}^{c}H \ , \label{opYuk-1} \\[1mm]
{\mathcal L}^{Maj} &\sim& \psi^{}_{}\phi_{}^{Maj}\psi^{}_{}HH
\ , \label{opMaj-1} \eea where both terms are invariant under
$G_{f}$, and we have suppressed flavour indices, order unity
coefficients and dimensional mass scales. Generically $\phi^{Yuk}$
and $\phi^{Maj}$ may represent either a single flavon or a
polynomial of flavons of a particular type. When the flavons
develop VEVs the family symmetry is spontaneously broken in such a
way that the full family symmetry is not apparent in either
of the low energy Lagrangians, only the observed flavour
symmetries corresponding to $S,U$ in ${\mathcal L}^{Maj}$
and $T$ in ${\mathcal L}^{Yuk}$.

\subsection{The flavons of direct models}
In direct models, as discussed in the Introduction, we should have
$G_f=S_4$ or a group that contains $S_4$ as a subgroup
(or $G_f=A_4$ as discussed below). In this approach one seeks to
identify the symmetry generators $S,U,T$ respected by ${\mathcal
L}^{Maj}$ and ${\mathcal L}^{Yuk}$, below the $G_f=S_4$
symmetry breaking scale, with the original generators contained in
$S_4$. One introduces three types of flavon denoted as $\phi_S$,
$\phi_U$, $\phi_T$, which each preserves a particular generator,
in other words the VEVs of these flavons are eigenvectors of
different generators of $G_f=S_4$ with eigenvalues of $+1$, \be
S\langle \phi_S \rangle =+1\langle \phi_S \rangle, \
\ U\langle \phi_U \rangle =+1 \langle \phi_U \rangle,
\ \  T\langle \phi_T
\rangle =+1\langle \phi_T \rangle, \label{flavons} \ee where
$\phi_{S,T}\sim {\bf 3}$ are in the triplet representation and $\phi_U
\sim {\bf 2}$ are in the doublet representation of $S_4$. With
Eq.\,(\ref{flavons}) the flavon VEVs $\langle \phi_S \rangle$ and $\langle
\phi_U \rangle$ are {\it both} left invariant under $S$ {\it as well as} under
$U$.\footnote{The generators of the doublet representation are
$S=\begin{pmatrix} 1&0\\0&1 \end{pmatrix}$, $U=\begin{pmatrix}
  0&1\\1&0 \end{pmatrix}$, $T=\begin{pmatrix}
  \omega&0\\0&\omega^2 \end{pmatrix}$. Therefore the VEV of the doublet
$\phi_U$ is trivially also an eigenvector of $S$ with eigenvalue
$+1$. Concerning the triplets in the basis of Eq.\,(\ref{SUgens}) the
eigenvector of $S$ with eigenvalue $+1$ is $\langle \phi_S \rangle \propto
\Phi_2$. In $S_4$ there are two distinct triplet representations which differ
in the sign of the $U$ generator in Eq.\,(\ref{SUgens}). In the case that
$\phi_S$ is in the triplet representation which corresponds to the positive
sign of $U$, then $\langle \phi_S \rangle \propto \Phi_2$ also preserves the
$U$ generator.}
In addition, the singlet flavon $\phi_I$ VEV preserves all the
generators. In the effective theory the flavons $\phi_{S,U}$ enter
the effective lepton operator linearly, so that $\phi^{Maj}$ is a
linear combination of $\phi_{S}$, $\phi_{U}$ and $\phi_I$, while
the flavon $\phi_T$ enters the Yukawa operators linearly, with
$\phi^{Yuk} \sim (\phi_T + \phi_I)$, \bea {\mathcal L}^{Yuk} &\sim&
\psi^{}_{}( \phi_{T}^{}+\phi_{I}^{})\psi_{}^{c}H \ , \label{opYuk-2} \\[1mm]
{\mathcal L}^{Maj} &\sim& \psi^{}_{}(\phi_{S}^{}+ \phi_{U}^{}
+ \phi_{I}^{}  )\psi^{}_{}HH, \label{opMaj-2} \eea where in the
Majorana Lagrangian $\psi \sim {\bf 3}$ represents lepton electroweak
doublets in the triplet representation of $S_4$, the Higgs
electroweak doublets $H \sim {\bf 1}$ are $S_4$ singlets and we have
suppressed flavour indices, order unity coefficients and
dimensional mass scales. As a consequence, in direct models, all
the flavon VEVs contained in $\phi^{Yuk}$ preserve the $T$
generator, while all the flavon VEVs contained in $\phi^{Maj}$
preserve the $S,U$ generators of the original family symmetry
$G_f$. Depending on the full symmetries of the model, $\phi_I$ may be replaced
by a pure number, leading to different suppressions between the terms in
the Yukawa and/or Majorana Lagrangian.

The type I see-saw mechanism \cite{Minkowski:1977sc}
in direct models exploits the flavons
$\phi_{S}^{},\phi_{U}^{},\phi_{I}^{}$ in Eq.\,(\ref{flavons}). In
general we could consider an $S_4$ model of the following form,
suppressing flavour indices, order unity coefficients and
dimensional mass scales, \bea {\mathcal L}^{Yuk}_N &\sim& L
(\phi_{S}^{}+ \phi_{U}^{} + \phi_{I}^{}  )      N^{c}H \ ,
\label{opYuk-5} \\[1mm] {\mathcal L}^{Maj}_{N} &\sim& N^{c}
(\phi_{S}^{}+ \phi_{U}^{} + \phi_{I}^{}  )    N^{c}\ ,
\label{opMaj-5} \eea
where $L$ represents lepton electroweak doublets, and
$N^c$ are the CP conjugated right-handed neutrinos.
Clearly the $S,U$ subgroups of $S_4$ are
preserved both in the neutrino Yukawa sector and the Majorana
sector. Then, after the see-saw mechanism, the same symmetries
$S,U$ must be apparent in the effective neutrino mass matrix.
Typically both the lepton doublets $L$ and the right-handed
neutrinos $N^c$ are taken to be triplets of $S_4$ or $A_4$.

The $A_4$ models in \cite{Altarelli:2006kg} provide a convenient
example of the application of $A_4$ to producing the neutrino
flavour symmetry in a direct way. These models are very well known
so here we only recall some of the salient features of these
models. The group $A_{4}$ \cite{Ma:2007wu} is a group that
describes even permutations of four objects. It has two
generators, $S$ and $T$, and four inequivalent irreducible
representations, ${\bf 1}$, ${\bf 1^{\prime}}$, ${\bf
1^{\prime\prime}}$ and ${\bf 3}$. The $S$ generator of the $A_4$
family symmetry survives in the Majorana sector and becomes part
of the neutrino flavour symmetry, while the $U$ transformation of
$S_4$ appears as an accidental symmetry of the neutrino mass
matrix arising from the absence of flavons in the ${\bf 1',1''}$
representations of $A_4$. Such $A_4$ examples satisfy Form
Dominance as discussed in \cite{Chen:2009um}, i.e. the neutrino
mass matrix may be diagonalised by the TB matrix independently of
the precise values of the underlying parameters. However in such
models each physical neutrino mass results from the VEVs of
different flavons, so some mild tuning of flavon VEVs is required
in order to achieve a neutrino mass hierarchy, as discussed in
\cite{Chen:2009um}.

\subsection{The flavons of indirect models}
In indirect models the family symmetry $G_f$ need not be
identified with $S_4$ or a group containing $S_4$ as a subgroup.
The idea is that the generators $S,U,T$ corresponding to the
flavour symmetries respected by ${\mathcal L}^{Maj}$ and
${\mathcal L}^{Yuk}$, below the $G_f$ symmetry breaking scale,
appear as accidental symmetries. Of course, the family symmetry
group $G_f$ will contain symmetry generators, but all elements obtained from
these generators will be broken completely by the flavon VEVs.
Nevertheless, the
combination of flavon VEVs appearing in $\phi^{Maj}$ will respect
the $S,U$ symmetry of the neutrino mass matrix, while the
combination of flavon VEVs appearing in $\phi^{Yuk}$ will respect
the $T$ symmetry of the charged lepton mass matrix (at least
approximately), even though neither $Z_2^S \times Z_2^U$ nor $Z_3^T$ are
subgroups of $G_f$.
In the cases where
$G_f$ contains some of the elements of $S_4$ what happens is
that all these elements will be broken by the flavon VEVs,
and new ones, analogous to the original ones but in a different
basis, will be accidentally restored in the low energy theory.

In indirect models one introduces triplet (or anti-triplet)
flavons of the family symmetry $G_f$, which we refer to as
$\phi_1$, $\phi_2$, $\phi_3$, which are arranged to get VEVs $v_i$
in the directions of the orthonormal column vectors in
Eq.\,(\ref{Phi0}), namely,
\be
\label{flavonvevs} \langle \phi_1
\rangle = v_1\Phi_1 , \ \ \langle \phi_2 \rangle =  v_2\Phi_2 , \
\ \langle \phi_3 \rangle =  v_3\Phi_3 .
\ee
The flavon VEV alignment $\phi_3$ was first discussed
in \cite{King:2003rf} and the analysis was extended to include
the flavon VEV alignment $\phi_2$ in \cite{King:2005bj,deMedeirosVarzielas:2005ax}.
The alignment of the flavons $\phi_2$, $\phi_3$ was subsequently discussed
in the framework of discrete family symmetries using F-terms in
\cite{deMedeirosVarzielas:2005qg} and using D-terms in
\cite{deMedeirosVarzielas:2006fc,King:2006np}.
The introduction of the flavon $\phi_1$ was
discussed in \cite{Chen:2009um}. The possibility that the three
flavons $\phi_i$ could be re-interpreted in terms
of string instantons was discussed in \cite{Antusch:2007jd},
although no instanton alignment mechanism was proposed.

In the effective
theory the flavons $\phi_{1,2,3}$ are arranged to enter the
effective lepton operator quadratically, so that $\phi^{Maj}$
consists of quadratic combinations of flavons combined as outer
products $\phi_{1}\phi_{1}^T$, $\phi_{2}\phi_{2}^T$ and
$\phi_{3}\phi_{3}^T$ so that the TB form of the neutrino mass
matrix in Eq.\,(\ref{eq:csd-tbm0}) is reproduced when the flavons get
their VEVs,
\be {\mathcal L}^{Maj} \sim
\psi^{}_{}(\phi_{1}\phi_{1}^T + \phi_{2}\phi_{2}^T +
\phi_{3}\phi_{3}^T)\psi^{}_{}HH, \label{opMaj-3} \ee
where $\psi \sim {\bf 3}$ represent lepton doublets in the real (complex)
triplet representation of $G_f$, while $\phi_i$ are triplets
(anti-triplets) of $G_f$, and we have suppressed flavour indices,
order unity coefficients and dimensional mass scales. The required
outer product structure of the quadratic flavons generally arises
from a see-saw mechanism, as discussed in section~5.

In indirect models the flavon VEVs in Eq.\,(\ref{flavonvevs}) break
all the elements of the underlying family symmetry $G_f$, while
the above quadratic combinations of VEVs preserve all the
group elements $G$ of the effective neutrino flavour symmetry, \be
\label{quad} G\langle \phi_{i}\phi_{i}^T \rangle G^T =\langle
\phi_{i}\phi_{i}^T \rangle, \ee for all $i=1,2,3$, which follows
from Eqs.~(\ref{S2}). In fact all the results of the previous section
apply here with the column vectors $\Phi_i$ replaced by the flavon
VEVs $\langle \phi_i \rangle$ in Eq.\,(\ref{flavonvevs}). In particular
from Eq.\,(\ref{G}), \be G \langle \phi_i \rangle = \eta_i \langle
\phi_i \rangle , \label{G3-new} \ee which shows that the flavon VEVs
individually break the flavour symmetry of the neutrino mass
matrix, corresponding to the group elements in Eq.\,(\ref{GG}), due to
the presence of the minus signs in the $\eta_i$, though their
quadratic effect is to preserve the neutrino flavour symmetry. For
example, for the group element $S=G_{-+-}= -G_1+G_2-G_3$, it is
clear from Eq.\,(\ref{GPhi}) that $S\langle \phi_1 \rangle =-\langle
\phi_1 \rangle$, $S\langle \phi_2 \rangle =\langle \phi_2
\rangle$, $S\langle \phi_3 \rangle =-\langle \phi_3 \rangle$, that
the VEVs $\langle \phi_1 \rangle$ and $\langle \phi_3 \rangle$
break the symmetry $S$.

We emphasise that, although the quadratic combinations of flavons
preserve an accidental neutrino flavour symmetry, such quadratic
combinations will in general break the underlying family symmetry
$G_f$. This does not matter as the only role of $G_f$ is to yield
flavon vacuum alignments of the type in Eq.\,(\ref{flavonvevs}).


\section{D-term vacuum alignment in indirect models}
The mechanism for vacuum alignment is crucial to the success of
any model which purports to explain tri-bimaximal mixing. In the
case of direct models, the usual mechanism of vacuum alignment is
based on F-term alignment which exploits driving fields in the
superpotential as discussed in \cite{Altarelli:2006kg}. This
mechanism is also available to indirect models as discussed in
\cite{deMedeirosVarzielas:2005qg}. However, in the case of
indirect models, an additional and elegant possibility for vacuum
alignment becomes available that is not possible for direct
models, namely D-term vacuum alignment introduced in
\cite{deMedeirosVarzielas:2006fc,King:2006np} as discussed in
section \ref{section-dtermpot}.

The reason that D-term vacuum alignment is not possible in the direct models
will be explained below, but is essentially related to the fact
that a particular choice of basis is required for the D-term
alignment mechanism to work, and this basis is different from that
of the neutrino flavour symmetry which must therefore arise
accidentally as in indirect models. In fact the D-term vacuum
alignment mechanism is so elegant that one may regard it as a
primary motivation for considering indirect models rather than
direct models.

\subsection[Flavon potential and allowed family symmetries
$G_f$]{\label{section-dtermpot}Flavon potential and allowed family symmetries $\boldsymbol{G_f}$}

In the previous section we emphasised that the flavons invoked in
indirect models generally break the family symmetry $G_f$
completely. An elegant way to obtain the alignments of
Eq.\,(\ref{Phi0}) is to start with a flavon scalar potential of
the form
\begin{equation}
V~=~
-m^2 \sum_i {\phi^i}^\dagger \phi^{i}_{}
~+~ \lambda \left(\sum_i {\phi^i}^\dagger \phi^{i}_{} \right)^2
~+~ \Delta V \ ,\label{nice-inv0}
\end{equation}
where the index $i$ labels the components of a particular flavon triplet
$\phi$ and
\begin{equation}
\Delta V ~=~\kappa \sum_i {\phi^i}^\dagger \phi^{i} {\phi^i}^\dagger \phi^{i}\ .
\label{nice-inv}
\end{equation}
In a non-supersymmetric theory this potential may simply be
written down. However, in a supersymmetric theory, the quartic
terms may arise from D-terms, after which this vacuum alignment
mechanism is named \cite{deMedeirosVarzielas:2006fc,King:2006np},
which take the form
$$
\left[ \chi^\dagger_{} \chi^{}_{} (\phi^\dagger_{} \phi^{}_{} \phi^\dagger_{}
\phi^{}_{} ) \right]_D \ ,
$$
where the F-component of the $G_f$ singlet $\chi$ acquires a VEV.
Hence supersymmetry is broken and the scalar potential $V$ gets a
contribution of the type $\phi^\dagger_{} \phi^{}_{}
\phi^\dagger_{} \phi^{}_{}$. The quadratic term on the other hand
can originate from a soft supersymmetry breaking mass term, where
the mass squared of a given flavon is driven negative by radiative
corrections at some scale $\Lambda$, leading to a VEV for that
flavon set by the scale $\Lambda$. As the different flavons have
different superpotential couplings to heavy states, and since the
soft masses run logarithmically with energy scale, the $\Lambda$
scales defined above may differ greatly for the different flavons.
Thus a hierarchy between the VEVs of various flavon fields is
possible, and also stable, in the framework of the radiative
breaking mechanism \cite{deMedeirosVarzielas:2006fc,King:2006np}.

Only the term $\Delta V$ in Eqs.(\ref{nice-inv0},\ref{nice-inv}) determines the alignment. For $\kappa>0$ we obtain
the alignment $\Phi_2$, while $\kappa<0$ can give rise to $\widetilde \Phi_0 =
(1,0,0)^T$. In concrete models, where both of these alignments are typically
required, the mixing terms between the two corresponding flavon fields should be
suppressed. Although such mixing cannot be forbidden by symmetries, it can be
suppressed or even forbidden by invoking messenger arguments.
Using $SU(3)$ invariant orthogonality conditions the alignments
$\Phi_3$ and $\Phi_1$ can then be obtained successively
\cite{deMedeirosVarzielas:2006fc,King:2009qt}, where additional undesirable
invariants may be suppressed or forbidden by a combination of symmetries and
messenger arguments. The question we
want to pursue in the following is which family symmetries allow for the
invariant in Eq.\,(\ref{nice-inv}). Obviously, $G_f$ must admit at least one
triplet representation.
Constraining to special unitary matrices, the
generators of the most general symmetry transformations of  Eq.\,(\ref{nice-inv})
are
\begin{equation}
D~=~\begin{pmatrix}
e^{i\theta_1} &0&0 \\
0& e^{i\theta_2}&0 \\
0&0& e^{-i(\theta_1+\theta_2)}
\end{pmatrix}  , \quad
A~=~\begin{pmatrix} 0&1&0\\0&0&1\\1&0&0\end{pmatrix}  , \quad
B~=~- \begin{pmatrix} 0&0&1 \\0&1&0 \\ 1&0&0\end{pmatrix}  .\label{gen-under}
\end{equation}
Note that these matrices are found among the generators of the C and D
series of groups listed long ago by Miller, Blichfeldt and Dickson \cite{MBD}.

Choosing $\theta_1=0$ and $\theta_2 = {2 \pi l}/{n}$ we immediately recover
the group $\Delta(6n^2)$ \cite{FFK,delta6n2}.
Dropping the generator $B$, we obtain the group $\Delta(3n^2)$
\cite{FFK,delta3n2} by identifying  $\theta_1 = {-2 \pi (k+l)}/{n}$ and
$\theta_2 = {2 \pi l}/{n}$. For these two series of non-Abelian finite groups,
$n,k,l$ are positive integers with $k,l<n$. A particular choice of $l$ (and $k$)
corresponds to a particular triplet representation of the respective group.
The Frobenius group $Z_7 \rtimes Z_3 =T_7$ \cite{Luhn:2007sy,Z7Z3} can be
generated from
$D$ and $A$ by setting $\theta_1 =\theta_2/2 = {2 \pi}/{7}$. Similar groups
that have not yet found application as family symmetries, although their order
is relatively low, are e.g.  $Z_{13} \rtimes Z_3$ with $\theta_1 =\theta_2/3 =
{2 \pi}/{13}$ or  $Z_{19} \rtimes Z_3$ with $\theta_1 =\theta_2/7 = {2
  \pi}/{19}$, see \cite{Fairbairn:1982jx}.

\subsection[The choice of basis: the $G_f=S_4$ example]{The choice of
basis: the $\boldsymbol{G_f=S_4}$ example}
It is easy to convince oneself that the neutrino flavour symmetry as given in
Eq.\,(\ref{GG}) is not a subgroup of any of the above finite groups.
In the special case of $\Delta_{24} \cong S_4$, which is generated by
\begin{equation}
 S'\,\equiv\, D_{(\theta_1=0,\theta_2=\pi)} \ , \qquad
 T'\,\equiv\, A \ , \qquad
 U'\,\equiv\, A^2 B 
  \ ,\label{S'T'}
\end{equation}
one might be tempted to think that, since $S_4$ is involved, that
the D-term alignment mechanism leads to an example of a direct model.
However this conclusion would be wrong, since it is clear that the generator
$S$ of the neutrino flavour symmetry in Eq.\,(\ref{SUgens}) is not an element
of the underlying group defined in the basis of Eq.\,(\ref{S'T'}).
In general such arguments, more fully developed below, will make it clear that the D-term alignment mechanism is incompatible
with the direct models, but well suited for the indirect models.

We now show that the choice of the basis in
Eq.\,(\ref{gen-under}) is crucial for having the invariant in Eq.\,(\ref{nice-inv}),
and thus for generating the alignment $\Phi_2$. Using a different basis for
$G_f$, the flavon potential $\Delta V$ would generally change its form
\begin{equation}
\sum_i {\phi^i}^\dagger \phi^{i} {\phi^i}^\dagger \phi^{i} ~~\longrightarrow ~~
\sum_{i,j,k,l,m} W^{}_{ji} W^\dagger_{ik} W^{}_{li} W^\dagger_{im} \
{{\phi^\prime}^j}^\dagger {\phi^{\prime}}^k {{\phi^\prime}^{l}}^\dagger {\phi^{\prime}}^{m} \ ,\label{nice-invNEW}
\end{equation}
where $W$ denotes the unitary basis transformation $\phi^\prime = W \phi$. In
this new form, the minima of $\Delta V$ are usually much less
apparent. However, since we already know how the flavon potential is minimised
in the original basis, we can trace back the structure of the vacuum for
$\phi^\prime$ to the alignment of $\phi$. It will turn out that no ``nice''
alignment can be obtained for $\phi^\prime$. To be specific, let us consider
the case $\kappa>0$. In the original basis, the flavon potential $\Delta V$
leads to an alignment
\begin{equation}
\widetilde \Phi_2 ~ = ~ \frac{1}{\sqrt{3}} \begin{pmatrix} e^{i \vartheta_1}
  \\ e^{i     \vartheta_2} \\ e^{i \vartheta_3} \end{pmatrix} \ .\label{alt-phi2}
\end{equation}
It is important to notice that the phases are completely undetermined by the
flavon potential of Eq.\,(\ref{nice-inv}). However these phases are unphysical
since they would correspond to having phases in the second column of $U_{TB}$
which can be absorbed into the charged lepton fields. What matters is the
magnitude of the components of Eq.\,(\ref{alt-phi2}) and the orthogonality of
the flavon VEVs $\widetilde \Phi_1,\widetilde \Phi_2,\widetilde \Phi_3$.
In indirect models it is a pure convention to set the phases of
Eq.\,(\ref{alt-phi2}) to zero, leading to $\Phi_2$. Nonetheless, the potential
$\Delta V$ of  Eq.\,(\ref{nice-inv}) is minimised by the more general
alignment of Eq.\,(\ref{alt-phi2}). Changing to the basis $\phi'$, an
analogous continuum of minima will persist, however, the explicit structure of
the set of alignment vectors  can appear arbitrary.

For example in $G_f=S_4$, suppose we perform a unitary basis transformation
\cite{Altarelli:2006kg}
\begin{equation}
W ~=~ \frac{1}{\sqrt{3}} \begin{pmatrix}
1 &1&1 \\ 1&\omega &\omega^2 \\  1&\omega^2 &\omega \end{pmatrix} \ , \qquad
\omega~=~e^{2\pi i/3} \ ,
\end{equation}
to take us from the basis of Eq.\,(\ref{S'T'}) to the neutrino flavour basis
of Eq.\,(\ref{SUgens}),
\begin{equation}
S\,=\,W S' W^\dagger \ , \quad
T\,=\,W T' W^\dagger \ , \quad
U\,=\,W U' W^\dagger \,=\,U'\ .
\end{equation}
Note that the third generator is identical in both bases. Nonetheless
  the neutrino flavour symmetry $Z_2^S\times Z_2^U$ is not a subgroup of $S_4$
generated by $S',T',U'$.
Performing the vacuum alignment in the basis of Eq.\,(\ref{S'T'}), would lead to
Eq.(\ref{alt-phi2}), which would appear in the neutrino flavour basis as,
\begin{equation}
\widetilde \Phi^\prime_2~=~W \widetilde \Phi_2 ~=~
\frac{1}{{3}} \begin{pmatrix}
e^{i \vartheta_1} + e^{i \vartheta_2} + e^{i \vartheta_3} \\
e^{i \vartheta_1} + \omega e^{i \vartheta_2} + \omega^2 e^{i \vartheta_3} \\
e^{i \vartheta_1} + \omega^2 e^{i \vartheta_2} + \omega e^{i \vartheta_3} \\
\end{pmatrix} .
\end{equation}
In general the original alignment would appear very arbitrary in the neutrino flavour basis, as can be seen from
the following examples,
$$
W \frac{1}{\sqrt{3}} \begin{pmatrix} 1\\1\\1  \end{pmatrix}
= \begin{pmatrix} 1\\0\\0  \end{pmatrix}, \quad
W \frac{1}{\sqrt{3}} \begin{pmatrix} 1\\-1\\-1  \end{pmatrix}
= \frac{1}{3} \begin{pmatrix} -1\\2\\2  \end{pmatrix}, \quad
W \frac{1}{\sqrt{3}} \begin{pmatrix} 1\\i\\-i  \end{pmatrix}
= \frac{1}{3} \begin{pmatrix} 1\\1-\sqrt{3}\\1+\sqrt{3}  \end{pmatrix} .
$$
These examples illustrate how dramatically the alignment vectors $\widetilde
\Phi^\prime_2$ in the new basis change with the phases $\vartheta_i$. Since
all of them minimise the flavon potential in the basis $\phi'$, the
invariant of Eq.\,(\ref{nice-invNEW}) does not seem to give
rise to useful alignments. That is not to say that a basis transformation
alters physics! Rather from the {\it practical} point of view the choice of a
good basis is important in devising an indirect model, starting from ``nice''
flavon alignments which are then coupled to the fermions to generate their
mass matrices.

To summarise, in the $S_4$ example with D-term vacuum alignment,
the above argument shows that the model
should be constructed in the $S',T',U'$ basis of Eq.\,(\ref{S'T'}) where the
alignment term in the potential is given by Eq.\,(\ref{nice-inv}) leading to the
vacuum alignment $\widetilde \Phi_2 $ in Eq.\,(\ref{alt-phi2}).
With the $G_f=S_4$ model constructed in the basis $S',T',U'$, the resulting
neutrino mass matrix would have a flavour symmetry corresponding to
the generators $S,U$. Due to the difference between the two bases the neutrino
flavour symmetry $Z_2^S\times Z_2^U$ is not a subgroup of the group generated by
$S',U',T'$ even though $U=U'$.
Furthermore, working in a convenient convention where the phases may be
dropped, i.e. $\widetilde \Phi_2 \rightarrow \Phi_2$,
the flavon alignments $\Phi_1,\Phi_2,\Phi_3$ guarantee that $S_4$
is broken completely: $\Phi_1$ breaks all elements of $S_4$, while $\Phi_2$
and $\Phi_3$ are left invariant by {\it distinct} generators,
\begin{equation}
T' \Phi_2~=~\Phi_2 \ , \qquad
U' \Phi_3~=~\Phi_3 \ .
\end{equation}
Having at least two flavons contributing to the neutrino sector, the
underlying family symmetry will therefore be broken completely. We conclude
that in the $G_f=S_4$ family symmetry model with D-term vacuum alignment, the
neutrino flavour symmetry arises accidentally, with the flavon alignment
$\Phi_2$ being directly enforced by the $G_f$ invariant term in
Eq.\,(\ref{nice-inv}). The alignment of the other flavons
$\Phi_1$, $\Phi_3$ of the indirect model arise from orthogonality arguments
and are not enforced directly from a $G_f$ invariant like
Eq.\,(\ref{nice-inv}), making their alignment more model dependent, see e.g.
\cite{King:2009qt}. This has
important  phenomenological implications, as discussed later in the context of
an indirect $G_f=A_4$ example where similar comments apply.

\subsection{Additional invariants non-grata}
Working in the basis of Eq.\,(\ref{gen-under}), we have seen that various
underlying non-Abelian discrete family symmetries are conceivable. They all
allow for the two quartic invariants of type ${\bf \overline 3 3 \overline 3
  3}$, where all triplets are assumed to be identical,
\begin{equation}
\mathcal I_0 ~=~ \left(\sum_i {\phi^i}^\dagger \phi^{i} \right)^2 \ , \qquad
\mathcal I_1 ~=~\sum_i {\phi^i}^\dagger \phi^{i} {\phi^i}^\dagger \phi^{i} \ .\label{inv0+1}
\end{equation}
Depending on $G_f$ there may however be additional invariants which we
identify in the following.

\begin{itemize}
\item $Z_7 \rtimes Z_3$: The symmetric Kronecker product of two triplets reads \cite{Luhn:2007sy,Z7Z3}
$$
({\bf 3} \times {\bf 3})_s ~=~{\bf 3} + {\bf \overline 3} \ .
$$
Multiplying this with its complex conjugate, i.e.
$$
({\bf 3} \times {\bf 3})_s \times ({\bf \overline 3} \times {\bf \overline
  3})_s ~=~
({\bf 3} + {\bf \overline  3}) \times ( {\bf \overline  3}+ {\bf 3}) \ ,
$$
it is evident that only two independent invariants of type ${\bf \overline 3 3
  \overline 3 3}$ are possible, namely $\mathcal I_0$ and $\mathcal I_1$.
The result is identical for the groups $Z_{13}\rtimes Z_3$ and
$Z_{19}\rtimes Z_3$ where $({\bf 3} \times {\bf 3})_s = {\bf 3'} +
{\bf \overline 3}$ with ${\bf 3'}$ different from ${\bf \overline
3}$.
\item $\Delta(3n^2)$: In this case the triplets are labelled
by two indices
  $k,l$. The symmetric product of two identical ${\bf 3}_{(k,l)}$ becomes
  \cite{delta3n2}
$$
({\bf 3}_{(k,l)} \times {\bf 3}_{(k,l)})_s ~=~ {\bf 3}_{(2k,2l)} + \,{\bf
  3}_{(-k,-l)} \ .
$$
Assuming that ${\bf 3}_{(k,l)}$ is a faithful representation of
$\Delta(3n^2)$, the two symmetric triplets on the right-hand side are
different triplets for $n>3$. Therefore, we find only two invariants in these
cases, $\mathcal I_0$ and $\mathcal I_1$.

With $n=2$ we recover $\Delta_{12} \cong A_4$. As there is only one real
triplet in $A_4$, the first symmetric triplet decomposes into a sum of one-dimensional representations,
$$
n=2:\quad ({\bf 3} \times {\bf 3})_s ~=~ {\bf 1}+{\bf 1'}+{\bf 1''} + {\bf 3} \ ,
$$
giving rise to four independent invariants. In addition to those of
Eq.\,(\ref{inv0+1}) we get \cite{King:2006np}
\begin{equation}
\mathcal I_2 ~=~
\phi^{1} \phi^{1} {\phi^2}^\dagger {\phi^2}^\dagger  +
\phi^{2} \phi^{2} {\phi^3}^\dagger {\phi^3}^\dagger  +
\phi^{3} \phi^{3} {\phi^1}^\dagger {\phi^1}^\dagger  \ ,
\end{equation}
and its complex conjugate $\mathcal I_{\bar 2} =\mathcal I_2^\dagger$.

With $n=3$ we obtain the group $\Delta_{27}$ with one triplet and its complex
conjugate. In this case, both symmetric triplets are identical,
$$
n=3:\quad ({\bf 3} \times {\bf 3})_s ~=~ {\bf \overline 3} + {\bf \overline 3} \ ,
$$
yielding four independent invariants. Besides $\mathcal I_0$ and
$\mathcal I_1$ which have been applied in \cite{deMedeirosVarzielas:2006fc} we
find (compare \cite{Luhn:2007sy})
\begin{equation}
\mathcal I_3 ~=~
\phi^{1} \phi^{1} {\phi^2}^\dagger {\phi^3}^\dagger  +
\phi^{2} \phi^{2} {\phi^3}^\dagger {\phi^1}^\dagger  +
\phi^{3} \phi^{3} {\phi^1}^\dagger {\phi^2}^\dagger  \ ,
\end{equation}
and its complex conjugate $\mathcal I_{\bar 3} =\mathcal I_3^\dagger$.
\item $\Delta(6n^2)$: For these groups there are two types of triplets
  ($q=1,2$) which differ by a sign for the generator $B$; each
  type of triplet is labelled by an additional index~$l$. The
  symmetric Kronecker products take the form \cite{delta6n2}
$$
({\bf 3_q}_{(l)} \times {\bf 3_q}_{(l)})_s ~=~ {\bf 3_1}_{(2l)} + \,{\bf 3_1}_{(-l)} \ .
$$
With ${\bf 3_q}_{(l)}$ being a faithful representation of $\Delta(6n^2)$, the
two triplets on the right-hand side are different triplets for $n>3$,  so that
only two quartic invariants are obtained, $\mathcal I_0$ and $\mathcal I_1$.

With $n=2$ we have $\Delta_{24} \cong S_4$, which has only two real
triplets. Furthermore, the first symmetric triplet decomposed into a
one-dimensional plus a two-dimensional representation,
$$
n=2: \quad ({\bf 3_q} \times {\bf 3_q})_s ~=~ {\bf 1} + {\bf 2} + {\bf 3_1} \ ,
$$ from which we can construct three independent invariants. In addition to
those of Eq.\,(\ref{inv0+1}) we find
\begin{equation}
\mathcal I_4 ~=~ \mathcal I_{2} + \mathcal I_{\bar 2}  \ .
\end{equation}

With $n=3$ the group $\Delta_{54}$ is generated, which has two complex triplets
and their conjugates. The symmetric Kronecker products are
$$
({\bf 3_q}_{(l)} \times {\bf 3_q}_{(l)})_s ~=~ {\bf 3_1}_{(-l)} + \,{\bf
  3_1}_{(-l)} \ ,
$$
leading to four independent invariants: $\mathcal I_0$, $\mathcal I_1$,
$\mathcal I_3$ and  $\mathcal I_{\bar 3}$. Note that these are identical to
the invariants of $\Delta_{27}$.
\end{itemize}

As we are interested in family symmetries which give rise to
``nice'' flavon alignments, it seems desirable to pick $G_f$ such
that only the invariants in Eq.\,(\ref{inv0+1}) are allowed.
Additional invariants may spoil the structure of the vacuum
derived from $\Delta V$ of Eq.\,(\ref{nice-inv}) unless they are
sufficiently suppressed. From that point of view, $\Delta(3n^2)$
and $\Delta(6n^2)$  with $n>3$, as well as $Z_7\rtimes Z_3$
together with some other examples such as $Z_{13}\rtimes Z_3$ and
$Z_{19}\rtimes Z_3$  are preferred candidates for the underlying
family symmetry in indirect models.

\subsection{Alternative useful invariants}
We conclude this section by mentioning that one could alternatively make use
of other invariants in the flavon potential that give rise to vacuum
alignments in indirect models. One such example could be the term
\begin{equation}
\mathcal I_5 ~=~ \phi^1 \phi^2\phi^3 \ , \label{nice-inv2}
\end{equation}
which is left invariant under $D$ and $A$ of Eq.\,(\ref{gen-under}). In other
words, the underlying discrete family symmetry $G_f$ in such a model could be
$Z_7\rtimes Z_3$ or $\Delta(3n^2)$. One easily finds that the flavon potential
with
$$
\Delta V = \kappa \,(\mathcal I^{}_5 + \mathcal I_5^\dagger) \ , \qquad
\kappa<0 \ ,
$$
is minimised by an alignment of type $\widetilde \Phi_2$,
$$
\frac{1}{\sqrt{3}} \begin{pmatrix}
e^{i \vartheta_1} \\  e^{i \vartheta_2} \\ e^{-i (\vartheta_1+\vartheta_2)}
\end{pmatrix} .
$$


\section{The see-saw mechanism in indirect models}
The type I see-saw mechanism \cite{Minkowski:1977sc}
in indirect models exploits the flavons
$\phi_i$ in Eq.\,(\ref{flavonvevs}) which are either triplets or
anti-triplets of $G_f$ depending on whether the representations
are complex. There are two approaches to the see-saw mechanism in
indirect models, depending on whether the right-handed neutrinos $N^c$ are
singlets or triplets under the family symmetry $G_f$, while the
left-handed leptons $L$ are always triplets of $G_f$. In both
cases we shall show how the quadratic combinations of flavons
preserve an accidental neutrino flavour symmetry of the neutrino
mass matrix, in the effective Lagrangian after the see-saw
mechanism has taken place.

\subsection[Right-handed neutrino singlets $N^c_i \sim 1$]{Right-handed
neutrino singlets $\boldsymbol{N^c_i \sim 1}$}
Consider the see-saw Lagrangian,
\bea {\mathcal L}^{Yuk}_N &\sim & L_i(
\phi_{1}^{i}N^{c}_1 + \phi_{2}^{i}N^{c}_2 + \phi_{3}^{i}N^{c}_3)H
\ , \label{opYuk-6} \\[1mm]
{\mathcal L}^{Maj}_{N} &\sim &M_{1}
N^{c}_1  N^{c}_1 +  M_{2}N^{c}_2  N^{c}_2 + M_{3}N^{c}_3   N^{c}_3
\ , \label{opMaj-6}
\eea
where the diagonal forms of
Eqs.~(\ref{opYuk-6},\ref{opMaj-6}) require additional symmetries.
Note that the Yukawa Lagrangian only involves the flavons $\phi_i$
linearly, not quadratically, and therefore does not respect the
flavour symmetry of the neutrino mass matrix which only emerges
after the see-saw mechanism. Since $N^c_i \sim {\bf 1}$, the combination
of $L_i \sim {\bf 3}$ and $\phi_i \sim {\bf 3}$ (or $\phi_i \sim
{\bf \overline{3}}$ if the representations are complex) must yield a
singlet of $G_f$. It is important to note
that these models are formulated in a basis where the
family indices are trivially summed over. After the see-saw
mechanism takes place, this results in an effective Lagrangian of
the form of Eq.\,(\ref{opMaj-3}),
\be {\mathcal L}^{Maj} \sim
L^{}_{} \left( \frac{\phi_{1}\phi_{1}^T}{M_1} +
\frac{\phi_{2}\phi_{2}^T}{M_2} +
\frac{\phi_{3}\phi_{3}^T}{M_3}\right) L^{}_{}HH.
\label{opMaj-7}
\ee
Thus we see the appearance of the quadratic
combinations of flavons which serve to preserve an accidental
neutrino flavour symmetry of the neutrino mass matrix, in the
effective Lagrangian after the see-saw mechanism has taken place.
In matrix notation, when the flavons get their VEVs in
Eq.\,(\ref{flavonvevs}) the three columns of the Dirac mass matrix
$M_D$ are proportional to the VEVs of the three flavons, and the
right-handed neutrino mass matrix is diagonal, \be M_D =
(a_1\Phi_1 , a_2\Phi_2 , a_3\Phi_3), \ \ M_{RR}= {\rm diag }(M_1,
M_2, M_3), \ee where $a_i$ are constants. The resulting effective
neutrino mass matrix is thus
\begin{equation}
M^{\nu} =  M_{D} M_{RR}^{-1} M_{D}^{T} =
 \frac{a_1^2}{M_1}\Phi_{1}\Phi_{1}^T + \frac{a_2^2}{M_2}\Phi_{2}\Phi_{2}^T +
\frac{a_3^2}{M_3}\Phi_{3}\Phi_{3}^T, \label{TBform}
\end{equation}
which is of TB form in Eq.\,(\ref{eq:csd-tbm0}) where we identify the
physical neutrino mass eigenvalues as $m_1=a_1^2/M_1$,
$m_2=a_2^2/M_2$, $m_3=a_3^2/M_3$. This also corresponds to Form
Dominance \cite{Chen:2009um}, with each column of $M_D$
constructed from the VEV of a different flavon, so no flavon VEV
tuning is required to achieve a neutrino mass hierarchy. This
corresponds to so-called Natural Form Dominance
\cite{Chen:2009um}.

The $A_4$ model in \cite{King:2006np} provides a convenient
example of the general mechanism described above for the realisation of neutrino
flavour symmetry in an indirect way using right-handed neutrino singlets. It also provides an example
of the use of CSD \cite{King:2005bj,King:1998jw}
to generate a strong neutrino mass hierarchy.
The lepton doublets are taken to be triplets of $A_4$, $L\sim {\bf 3}$,
while the right-handed neutrinos and right-handed charged leptons
are all taken to be trivial singlets of $A_4$. In this model the
see-saw Lagrangian is taken to be, at leading order, similar to
the form in Eqs.~(\ref{opYuk-6},\ref{opMaj-6}), but with some of the
right-handed neutrinos re-labelled, and involves the $A_4$ triplet
flavons $\phi_2^i \sim {\bf 3}$, $\phi_3^i \sim {\bf 3}$ plus a new triplet
flavon $\phi_0^i \sim {\bf 3}$,
\bea {\mathcal L}^{Yuk}_N &\sim&
L_i( \phi_{3}^{i}N^{c}_1 + \phi_{2}^{i}N^{c}_2 + \phi_{0}^{i}N^{c}_3)H
\ , \label{opYuk-10} \\[1mm] {\mathcal L}^{Maj}_{N} &\sim & M_{1}
N^{c}_1  N^{c}_1 +  M_{2}N^{c}_2  N^{c}_2 + M_{3}N^{c}_3   N^{c}_3
\ , \label{opMaj-10}
\eea
where the assumed forms in
Eqs.~(\ref{opYuk-10},\ref{opMaj-10}) require additional symmetries as
discussed in \cite{King:2006np}. The flavons $\phi_2$, $\phi_3$
have the alignments as in Eq.\,(\ref{flavonvevs}) while the flavon
$\phi_0$ has an alignment of the form,
\begin{equation}
\label{A4vevs3} \left< \phi_{0} \right> = \left( \begin{array}{c}
0 \\ 0 \\ 1 \end{array} \right)v_0 \equiv \Phi_0 v_0\; .
\end{equation}
The charged lepton Yukawa Lagrangian also takes a similar form to
Eq.\,(\ref{opYuk-10}), with the flavon $\phi_0$ being responsible for
the third family charged lepton Yukawa coupling,
\be
{\mathcal
L}^{Yuk}_{\mbox{\scriptsize lep}} \sim L_i( \phi_{3}^{i}e^{c}_R +
\phi_{2}^{i}\mu^{c}_R + \phi_{0}^{i}\tau^{c}_R)H      \ .
\label{opYuk-11}
\ee
The resulting charged lepton mass matrix is
approximately diagonal due to the strong charged lepton mass
hierarchy with the dominant $(3,3)$ Yukawa coupling provided by
the $\phi_0$ VEV. As pointed out in section \ref{section-dtermpot}, the required
hierarchy between different flavon VEVs can be obtained in the framework of
radiative symmetry breaking. The actual model in \cite{King:2006np} is more
complicated than this, since it unifies the quarks and leptons
into a Pati-Salam gauge group, under which for example the
right-handed neutrinos are not singlets, and shows how all the
fermion mass hierarchies and mass splittings may be achieved.

In this $A_4$ model the neutrino flavour symmetry arises accidentally as
a result of two effects. The first effect is CSD \cite{King:2005bj,King:1998jw}
in which the right-handed neutrino mass $M_3$ is so heavy\footnote{In the
concrete model \cite{King:2006np}, additional symmetries are adopted so that
the third term on the RHS of Eq.\,(\ref{opMaj-10}) arises at a
lower-dimensional level compared to the first and second term.} that the third
right-handed neutrino $N^c_3$ effectively decouples from the
see-saw mechanism, rendering the effect of the flavon $\phi_0$ on
the see-saw mechanism irrelevant \cite{King:1998jw}. The second
effect is that the remaining two flavons $\phi_2$, $\phi_3$ which
are relevant for the see-saw mechanism lead to an effective
Majorana Lagrangian after the see-saw mechanism of the form of
Eq.\,(\ref{opMaj-7}),
\be
{\mathcal L}^{Maj} \sim L
\left(\frac{\phi_{3}\phi_{3}^T}{M_1}+\frac{\phi_{2}\phi_{2}^T}{M_2}\right)
L HH. \label{opMaj-12}
\ee
Thus we see the appearance of the
quadratic combinations of flavons which serve to preserve an
accidental neutrino flavour symmetry of the neutrino mass matrix,
in the effective Lagrangian after the see-saw mechanism has taken
place. However, since the charged lepton mass matrix is only
approximately diagonal, TB neutrino mixing will receive
corrections from the charged lepton mixing angles
\cite{Antusch:2008yc}.

An important feature of the indirect $A_4$ model is that it is
formulated in an $SO(3)$ type basis for which the product of two
triplets $a=(a_1,a_2,a_3)\sim {\bf 3}$ and $b=(b_1,b_2,b_3)\sim
{\bf 3}$ contains the invariant singlet given by the diagonal
combination $a_ib_i=a_1b_1+a_2b_2+a_3b_3 \sim {\bf 1}$. In the
$A_4$ example \cite{King:2006np} this implies that the generators
in the triplet representation are given by $S', T'$ of
Eq.\,(\ref{S'T'}), see also \cite{Ma:2007wu}. On the other hand
these models reproduce the neutrino flavour symmetry $Z_2^S\times
Z_2^U$, which however, is not a subgroup of the original $A_4$
family symmetry. The reason for this is twofold: ($i$) the
generator $U$ is not an element of $A_4$, and ($ii$) the bases are
different. Furthermore, the VEV of $\phi_3$ breaks all elements
generated from $S',T'$, while $\phi_2$ preserves the subgroup
$Z_3^{T'}$, the net effect being that no subgroup of the
underlying $A_4$ symmetry can survive in either the neutrino or
the charged lepton sector.

Contrary to the $\phi_2$ alignment which is obtained directly from
the invariant in Eq.\,(\ref{nice-inv}), the $\phi_3$ alignment
emerges from subsequent orthogonality arguments and is thus more
model dependent, see e.g. \cite{King:2009qt}. While the alignment of $\phi_2$
may be altered by subleading correction, it is possible for the VEV of
$\phi_3$ to receive a {\it leading order} alignment of the form
$(\epsilon,1,-1)$, where the value of $\epsilon$ is potentially sizable
\cite{King:2009qt}. This would then result in a reactor angle of order
$\theta_{13}\sim \epsilon / \sqrt{2}$, at the same time maintaining accurately
the tri-bimaximal solar and atmospheric predictions
\cite{King:2009qt}. Such tri-bimaximal-reactor mixing of the type
discussed in \cite{King:2009qt} would then be a prediction of
indirect models with such a leading order $\phi_3$ alignment.

\subsection[Right-handed neutrino triplets $N^c_i \sim 3$]{Right-handed
neutrino triplets $\boldsymbol{N^c_i \sim 3}$}
In this case the see-saw Lagrangian is taken to have the form, at
leading order,
\bea {\mathcal L}^{Yuk}_N &\sim &  L_i \left(
\phi_{1}^i\phi_{1}^{j} + \phi_{2}^i\phi_{2}^j + \phi_{3}^i\phi_{3}^j
\right) N^{c}_j  H     \ , \label{opYuk-8} \\[1mm]
 {\mathcal L}^{Maj}_{N} &\sim & N^c_i \left( \phi_{1}^i\phi_{1}^j +
\phi_{2}^i\phi_{2}^j + \phi_{3}^i\phi_{3}^j\right) N^c_j   \ ,
\label{opMaj-8}
\eea
where the Yukawa and Majorana Lagrangians are
required to be diagonal in the flavon types, up to a relabelling
of right-handed neutrinos. Note that in this case both the Yukawa
and Majorana Lagrangian involve the flavons $\phi_i$
quadratically, and therefore in this case the high energy see-saw
theory respects the same flavour symmetry of the neutrino mass
matrix which will emerge after the see-saw mechanism, although of
course the family symmetry $G_f$ is not respected by these
quadratic flavon combinations.

As before we see the appearance of the
quadratic combinations of flavons which serve to preserve an
accidental neutrino flavour symmetry of the neutrino mass matrix,
in the effective Lagrangian after the see-saw mechanism has taken
place. In matrix notation, when the flavons get their VEVs in
Eq.\,(\ref{flavonvevs}), the Dirac mass matrix $M_D$ is proportional
to a sum over outer products of the VEVs of the three flavons,
while the right-handed neutrino mass matrix takes the form of the
TB neutrino mass matrix in Eq.\,(\ref{eq:csd-tbm0}),
\bea M_D &=&  a_{1}
\Phi_{1}\Phi_{1}^{T} + a_{2} \Phi_{2}\Phi_{2}^{T} + a_{3}
\Phi_{3}\Phi_{3}^{T}   \ , \label{matrices0} \\[1mm]
 M_{RR}&=& M_{1}
\Phi_{1}\Phi_{1}^{T} + M_{2} \Phi_{2}\Phi_{2}^{T} + M_{3}
\Phi_{3}\Phi_{3}^{T}   \ , \label{matrices}
\eea
where $a_i$ are
constants and $M_i$ are the right-handed neutrino mass
eigenvalues. Clearly the right-handed neutrino mass matrix is
diagonalised by the tri-bimaximal mixing matrix,
\begin{equation}
U_{TB}^{T} {M_{RR}} U_{TB}=\mathrm{diag}\,(M_{1}, \; M_{2}, \;
M_{3}) \; .
\end{equation}
Hence we can write
\begin{equation}
{M_{RR}^{-1}} = \frac{\Phi_{1}\Phi_{1}^T}{M_1} +
\frac{\Phi_{2}\Phi_{2}^T}{M_2} + \frac{\Phi_{3}\Phi_{3}^T}{M_3}\;,
\label{inverse}
\end{equation}
using Eq.\,(\ref{columns}). The resulting effective neutrino mass
matrix is thus, from Eqs.~(\ref{matrices0},\ref{inverse}), using the
orthonormality relations in Eq.\,(\ref{orthonormal}),
\begin{equation}
M^{\nu} =  M_{D} M_{RR}^{-1} M_{D}^{T} =
\frac{a_1^2}{M_1}\Phi_{1}\Phi_{1}^T +
\frac{a_2^2}{M_2}\Phi_{2}\Phi_{2}^T +
\frac{a_3^2}{M_3}\Phi_{3}\Phi_{3}^T, \label{meff}
\end{equation}
as in Eq.\,(\ref{eq:csd-tbm0}), where we identify the physical neutrino
mass eigenvalues as $m_1=a_1^2/M_1$, $m_2=a_2^2/M_2$,
$m_3=a_3^2/M_3$.

The $G_f=\Delta_{27}$ model in \cite{deMedeirosVarzielas:2006fc} provides an
example of the general mechanism described above for the realisation of neutrino
flavour symmetry in an indirect way using right-handed neutrino triplets. It also provides an example
of the use of CSD \cite{King:2005bj,King:1998jw}
to generate a strong neutrino mass hierarchy $m_1 \ll m_{2,3}$
due to a very heavy third right-handed neutrino of mass $M_3$.
The Yukawa Lagrangian is of the leading order form,
\be {\mathcal
L}^{Yuk}_N \sim   L_i \left( \phi_{3}^i\phi_{2}^j +
\phi_{2}^i\phi_{3}^j + \phi_{0}^{i}\phi_{0}^{j} \right) N^{c}_j  H     \ , \label{opYuk-9}
\ee
leading to a Dirac mass matrix of the form,
\be M_D = a_{1} \Phi_{3}\Phi_{2}^{T} +  a_{2} \Phi_{2}\Phi_{3}^{T}  + a_{0} \Phi_{0}\Phi_{0}^{T}, \ \
\label{matrices2}
\ee
where $a_i$ are
constants.
The Majorana Lagrangian is of the leading order form,
\be {\mathcal L}^{Maj}_{N}
\sim N^c_i \left( \phi_{2}^i\phi_{2}^j  +
\phi_{3}^i\phi_{3}^j + \phi_{0}^i\phi_{0}^j \right) N^c_j   \ ,
\label{opMaj-22}
\ee
leading to the heavy Majorana masses,
\be M_{RR}=
M_{1} \Phi_{2}\Phi_{2}^{T} +
M_{2} \Phi_{3}\Phi_{3}^{T} +
M_{3} \Phi_{0}\Phi_{0}^{T}    \ , \label{matrices6}
\ee
where $M_i$ are the right-handed neutrino mass
eigenvalues where we assume $M_1 < M_2 \ll M_3$.
Assuming that the heaviest right-handed neutrino of mass $M_3$ approximately decouples,
according to the CSD mechanism,
the resulting effective neutrino mass
matrix is thus, from Eqs.~(\ref{matrices2},\ref{matrices6}), using the
orthonormality relations in Eq.\,(\ref{orthonormal}),
\begin{equation}
M^{\nu} =  M_{D} M_{RR}^{-1} M_{D}^{T} \approx
\frac{a_1^2}{M_1}\Phi_{3}\Phi_{3}^T +
\frac{a_2^2}{M_2}\Phi_{2}\Phi_{2}^T, \label{meff-new}
\end{equation}
similar to Eq.\,(\ref{eq:csd-tbm0}),
where $m_1 \ll m_{2,3}$ is negligible
and the remaining physical neutrino mass eigenvalues are
$m_3=a_1^2/M_1$, $m_2=a_2^2/M_2$. Note that in this case the
Yukawa Lagrangian involves off-diagonal flavon bilinears
$\phi_i\phi_j$ (i.e. two different flavon types)
and therefore it does not respect the flavour symmetry of
the neutrino mass matrix which only emerges after the see-saw
mechanism.


\section{Conclusion}
In this paper we have shown that the flavour symmetry of the
neutrino mass matrix may originate from an underlying family
symmetry in two quite distinct ways, either directly or
indirectly.

The direct models are typically based on $S_4$ family symmetry or
any family symmetry that contains $S_4$ as a subgroup
(such as for example $PSL(2,7) = \Sigma (168)$ \cite{King:2009mk})
whose generators $S,U$ are directly preserved in
the Majorana sector sector.
In the case of direct models the neutrino flavour
symmetry appears directly as a result of a $Z_2^S \times Z_2^U$
subgroup of $S_4$ being preserved in the effective Majorana
Lagrangian by the flavons $\phi_{S}^{}, \phi_{U}^{} $, but not $\phi_{T}^{}$,
appearing in the Majorana sector.
In the see-saw implementation of the direct models the neutrino
flavour symmetry is preserved also by the high energy see-saw
theory, due to the presence of the flavons $\phi_{S}^{},
\phi_{U}^{}$, but not $\phi_{T}^{}$, appearing in the see-saw
Dirac and Majorana Lagrangians as in
Eqs.~(\ref{opYuk-5},\ref{opMaj-5}).

The main focus of the paper has been on indirect models which are based on
any family symmetry $G_f$ which is completely broken and in which the
observed lepton flavour symmetry emerges as an accidental
symmetry. In the case of indirect models the only role of the underlying family symmetry
$G_f$ is to yield flavon alignments with $\phi_i$ proportional
to the three columns of $U_{TB}$ as in
Eq.\,(\ref{flavonvevs}). The flavon VEVs of indirect models
break the underlying family symmetry $G_f$, so that no
remnant of this symmetry survives in the low energy theory.
The origin of the neutrino flavour symmetry is completely
accidental due to the quadratic appearance of such flavons in
the effective Majorana Lagrangian.

Comparing the direct models to the indirect models it is seen that,
for the direct models, the flavons
$\phi_{S}^{}, \phi_{U}^{} $ appear in the Majorana sector as in
Eq.\,(\ref{opMaj-2}), while for the indirect models it is the
quadratic flavon combinations
$\phi_i \phi_i^T$ which appear in the Majorana sector as in
Eq.\,(\ref{opMaj-3}), where such quadratic combinations accidentally
preserve the neutrino flavour symmetry and reproduce the TB
neutrino mass matrix.

We have emphasised that the necessary vacuum alignments of indirect models can be
achieved using an elegant D-term vacuum alignment mechanism, together with orthogonality arguments,
and we have catalogued the possible
choices of family symmetry $G_f$ which are consistent with this mechanism.
In this way we are led to the large classes of possible candidate family
symmetries:
$\Delta(3n^2)$ and $\Delta(6n^2)$, as well as $Z_7 \rtimes Z_3 =T_7$
together with similar examples such as
$Z_{13} \rtimes Z_3$ and $Z_{19} \rtimes Z_3$ which so far have been ignored as
potentially interesting candidates for $G_f$.
Although the presence of the underlying
family symmetry $G_f$ is crucial for producing such D-term vacuum alignments,
we have shown that it does not include the neutrino flavour symmetry
$Z_2^S\times Z_2^U$ as a subgroup which must therefore emerge as an accidental
symmetry.
We have explicitly shown how this works
for the case of $G_f=S_4$ where D-term vacuum alignment implies an indirect, rather than direct,
realisation of the neutrino flavour symmetry.

We have seen that the see-saw implementation of the indirect models
depends on whether the right-handed neutrinos are singlets or triplets of the
family symmetry $G_f$ and may be summarised as follows:

(i) If the right-handed neutrinos are singlets,
then the flavons $\phi_i$ appear linearly in the Dirac Lagrangian,
as in Eq.\,(\ref{opYuk-6}), so that the high energy see-saw theory
does not respect the low energy neutrino flavour symmetry of the
resulting low energy effective theory in Eq.\,(\ref{opMaj-7}) which
only involves the quadratic flavon combinations.

(ii) If the right-handed neutrinos are triplets, then the quadratic flavon
combinations $\phi_i \phi_i^T$ may appear in both the high energy
Dirac and Majorana Lagrangians, as in
Eqs.~(\ref{opYuk-8},\ref{opMaj-8}), which means that the high energy
see-saw theory will respect the low energy neutrino flavour
symmetry in such cases. However in other cases of indirect models
in which the right-handed neutrinos are triplets, the right-handed
neutrinos may be ordered differently so that an off-diagonal
flavon combination $\phi_i \phi_j^T$ may appear in the high energy
Dirac Lagrangian, as in Eq.\,(\ref{opYuk-9}), with the diagonal
combinations $\phi_i \phi_i^T$ only emerging in the low energy
effective Majorana Lagrangian so that the neutrino flavour
symmetry only arises below the see-saw scale.

We have seen that explicit realistic see-saw implementations of
indirect models are typically based on CSD which implies a strong
neutrino hierarchy and involves only the two flavons with
alignments along the second and third columns of the TB mixing
matrix. We have sketched existing examples of such models based on
$\Delta_{12}=A_4$ and $\Delta_{27}$.

In conclusion, the main point we want to make in this paper is
that the family symmetry group $G_f$ need not have anything
directly to do with the observed neutrino flavour symmetry. In
such indirect models the only role of $G_f$ is to produce the
vacuum alignments proportional to the columns of $U_{TB}$, as in
Eq.\,(\ref{flavonvevs}), where the quadratic appearance of such
flavons in the effective Majorana Lagrangian is responsible for
the observed neutrino flavour symmetry corresponding to TB mixing.
In such indirect models, D-term vacuum alignment provides an
elegant vacuum alignment mechanism, not available to the direct
models. In fact the D-term vacuum alignment mechanism is so
elegant that one may regard it as a primary motivation for
considering indirect models rather than direct models. The D-term
vacuum alignment mechanism only requires $G_f$ to contain triplet
representations of the form of Eq.\,(\ref{gen-under}) which is the
case for the large classes of finite groups mentioned above.
Typically the VEV of the flavon $\phi_2$, proportional
  to the second column of $U_{TB}$, will be
  derived directly from the $G_f$ invariant term in Eq.\,(\ref{nice-inv}),
  whereas the alignments of the other flavons such as $\phi_3$, proportional to the
  third column of $U_{TB}$, are
  subsequently obtained using orthogonality arguments. In such indirect models
  the alignment of $\phi_3$ is therefore generally more model dependent
and, for example, could lead to a large reactor angle while
preserving the TB solar and atmospheric angle predictions
\cite{King:2009qt}, providing a smoking gun signature of the
indirect origin of the neutrino flavour symmetry.


\section*{Acknowledgements}
We acknowledge partial support from the STFC
Rolling Grant ST/G000557/1.



\begin{thebibliography}{99}

\bibitem{HPS}
P.~F.~Harrison, D.~H.~Perkins and W.~G.~Scott,
{\em Tri-bimaximal mixing and the neutrino oscillation data,}
Phys.\ Lett.\ B {\bf 530} (2002) 167 [hep-ph/0202074].

\bibitem{Harrison:2003aw}
P.~F.~Harrison and W.~G.~Scott,
{\em Permutation symmetry, tri-bimaximal neutrino mixing and the $S_3$ group
characters,}
Phys.\ Lett.\ B {\bf 557} (2003) 76 [hep-ph/0302025].

\bibitem{Ma:2007wu}
  E.~Ma and G.~Rajasekaran,
  {\em Softly broken $A_4$ symmetry for nearly degenerate neutrino masses,}
  Phys.\ Rev.\  D {\bf 64} (2001) 113012
  [hep-ph/0106291].

\bibitem{King:2005bj}
  S.~F.~King,
  {\em Predicting neutrino parameters from SO(3) family symmetry and
    quark-lepton unification,}
  JHEP {\bf 0508}, 105 (2005)
  [hep-ph/0506297].

\bibitem{deMedeirosVarzielas:2005ax}
  I.~de Medeiros Varzielas and G.~G.~Ross,
  {\em SU(3) family symmetry and neutrino bi-tri-maximal mixing,}
  Nucl.\ Phys.\  B {\bf 733} (2006) 31
  [hep-ph/0507176].

\bibitem{King:2006me}
  S.~F.~King and M.~Malinsky,
  {\em Towards a complete theory of fermion masses and mixings with SO(3)
    family symmetry and 5d SO(10) unification,}
  JHEP {\bf 0611} (2006) 071
  [hep-ph/0608021].

\bibitem{deMedeirosVarzielas:2005qg}
  I.~de Medeiros Varzielas, S.~F.~King and G.~G.~Ross,
  {\em Tri-bimaximal neutrino mixing from discrete subgroups of SU(3) and  SO(3)
  family symmetry,}
  Phys.\ Lett.\  B {\bf 644} (2007) 153
  [hep-ph/0512313].

\bibitem{deMedeirosVarzielas:2006fc}
  I.~de Medeiros Varzielas, S.~F.~King and G.~G.~Ross,
  {\em Neutrino tri-bi-maximal mixing from a non-Abelian discrete family
  symmetry,}
  Phys.\ Lett.\  B {\bf 648} (2007) 201
  [hep-ph/0607045].

\bibitem{King:2006np}
  S.~F.~King and M.~Malinsky,
  {\em $A_4$ family symmetry and quark-lepton unification,}
  Phys.\ Lett.\  B {\bf 645} (2007) 351
  [hep-ph/0610250].

\bibitem{Altarelli:2006kg}
  G.~Altarelli,
  {\em Models of neutrino masses and mixings,}
  hep-ph/0611117;\\
%
  G.~Altarelli, F.~Feruglio and Y.~Lin,
  {\em Tri-bimaximal neutrino mixing from orbifolding,}
  Nucl.\ Phys.\  B {\bf 775} (2007) 31
  [hep-ph/0610165];\\
%
  G.~Altarelli and F.~Feruglio,
  {\em Tri-bimaximal neutrino mixing, $A_4$ and the modular symmetry,}
  Nucl.\ Phys.\  B {\bf 741} (2006) 215
  [hep-ph/0512103];\\
%
  G.~Altarelli and F.~Feruglio,
  {\em Tri-bimaximal neutrino mixing from discrete symmetry in extra
  dimensions,}
  Nucl.\ Phys.\  B {\bf 720} (2005) 64
  [hep-ph/0504165].

\bibitem{Chen:2009um}
  M.~C.~Chen and S.~F.~King,
  {\em $A_4$ see-saw models and form dominance,}
  JHEP {\bf 0906} (2009) 072
  [arXiv:0903.0125].

\bibitem{Frampton:2004ud}
  P.~H.~Frampton, S.~T.~Petcov and W.~Rodejohann,
  {\em On deviations from bimaximal neutrino mixing,}
  Nucl.\ Phys.\  B {\bf 687} (2004) 31
  [hep-ph/0401206];\\
%
  F.~Plentinger and W.~Rodejohann,
  {\em Deviations from tri-bimaximal neutrino mixing,}
  Phys.\ Lett.\ B {\bf 625} (2005) 264
  [hep-ph/0507143];\\
%
  R.~N.~Mohapatra and W.~Rodejohann,
  {\em Broken mu-tau symmetry and leptonic CP violation,}
  Phys.\ Rev.\ D {\bf 72} (2005) 053001
  [hep-ph/0507312];\\
%
  K.~A.~Hochmuth, S.~T.~Petcov and W.~Rodejohann,
  {\em $U_{PMNS} = U_\ell^\dagger U_\nu$,}
  arXiv:0706.2975;\\
%
  T.~Ohlsson and G.~Seidl,
  {\em A flavor symmetry model for bilarge leptonic mixing and the lepton
  masses,}
  Nucl.\ Phys.\  B {\bf 643} (2002) 247
  [hep-ph/0206087];\\
%
  E.~Ma,
  {\em Near tri-bimaximal neutrino mixing with $\Delta(27)$ symmetry,}
  arXiv:0709.0507;\\
%
  E.~Ma,
  {\em New lepton family symmetry and neutrino tribimaximal mixing,}
  hep-ph/0701016;\\
%
  E.~Ma,
  {\em Supersymmetric $A_4 \times Z_3$ and $A_4$ realisations of neutrino
    tribimaximal mixing without and with corrections,}
  Mod.\ Phys.\ Lett.\  A {\bf 22} (2007) 101
  [hep-ph/0610342];\\
%
  E.~Ma,
  {\em Suitability of $A_4$ as a family symmetry in grand unification,}
  Mod.\ Phys.\ Lett.\  A {\bf 21} (2006) 2931
  [hep-ph/0607190];\\
%
  E.~Ma,
  {\em Neutrino mass matrix from $\Delta(27)$ symmetry,}
  Mod.\ Phys.\ Lett.\  A {\bf 21} (2006) 1917
  [hep-ph/0607056];\\
%
  E.~Ma, H.~Sawanaka and M.~Tanimoto,
  {\em Quark masses and mixing with $A_4$ family symmetry,}
  Phys.\ Lett.\  B {\bf 641} (2006) 301
  [hep-ph/0606103];\\
%
  E.~Ma,
  {\em Tri-bimaximal neutrino mixing from a supersymmetric model with $A_4$
    family symmetry,}
  Phys.\ Rev.\  D {\bf 73} (2006) 057304
  [hep-ph/0511133];\\
%
  B.~Adhikary, B.~Brahmachari, A.~Ghosal, E.~Ma and M.~K.~Parida,
  {\em $A_4$ symmetry and prediction of $U_{e3}$ in a modified
    Altarelli-Feruglio model,}
  Phys.\ Lett.\  B {\bf 638} (2006) 345
  [hep-ph/0603059];\\
%
  E.~Ma,
  {\em Tetrahedral family symmetry and the neutrino mixing matrix,}
  Mod.\ Phys.\ Lett.\  A {\bf 20} (2005) 2601
  [hep-ph/0508099];\\
%
  E.~Ma,
  {\em Aspects of the tetrahedral neutrino mass matrix,}
  Phys.\ Rev.\  D {\bf 72} (2005) 037301
  [hep-ph/0505209];\\
%
  S.~L.~Chen, M.~Frigerio and E.~Ma,
  {\em Hybrid see-saw neutrino masses with $A_4$ family symmetry,}
  Nucl.\ Phys.\  B {\bf 724} (2005) 423
  [hep-ph/0504181];\\
%
  E.~Ma,
  {\em $A_4$ origin of the neutrino mass matrix,}
  Phys.\ Rev.\  D {\bf 70} (2004) 031901
  [hep-ph/0404199];\\
%
  F.~Feruglio, C.~Hagedorn, Y.~Lin and L.~Merlo,
  {\em Tri-bimaximal neutrino mixing and quark masses from a discrete flavour
  symmetry,}
  Nucl.\ Phys.\  B {\bf 775} (2007) 120
  [hep-ph/0702194];\\
%
  F.~Plentinger and G.~Seidl,
  {\em Mapping out SU(5) GUTs with non-Abelian discrete flavor symmetries,}
  Phys.\ Rev.\  D {\bf 78} (2008) 045004
  [arXiv:0803.2889];\\
%
  C.~Csaki, C.~Delaunay, C.~Grojean and Y.~Grossman,
  {\em A model of lepton masses from a warped extra dimension,}
  arXiv:0806.0356;\\
%
  M.-C.~Chen and K.~T.~Mahanthappa,
  {\em CKM and tri-bimaximal MNS matrices in a $SU(5)\times\,^{d}T$ model,}
  Phys.\ Lett.\  B {\bf 652} (2007) 34
  [arXiv:0705.0714];\\
%
  M.-C.~Chen and K.~T.~Mahanthappa,
  {\em Tri-bimaximal neutrino mixing and CKM matrix in a $SU(5)\times\,^{d}T$
    model,}
  arXiv:0710.2118;\\
%
  M.-C.~Chen and K.~T.~Mahanthappa,
  {\em Neutrino mass models: circa 2008,}
  arXiv:0812.4981;\\
%
  R.~N.~Mohapatra, S.~Nasri and H.~B.~Yu,
  {\em $S_3$ symmetry and tri-bimaximal mixing,}
  Phys.\ Lett.\  B {\bf 639} (2006) 318
  [hep-ph/0605020];\\
%
  R.~N.~Mohapatra and H.~B.~Yu,
  {\em Connecting leptogenesis to CP violation in neutrino mixings in a
  tri-bimaximal mixing model,}
  Phys.\ Lett.\  B {\bf 644} (2007) 346
  [hep-ph/0610023];\\
%
  X.~G.~He,
  {\em $A_4$ group and tri-bimaximal neutrino mixing: a renormalizable model,}
  Nucl.\ Phys.\ Proc.\ Suppl.\  {\bf 168} (2007) 350
  [hep-ph/0612080];\\
%
  A.~Aranda,
  {\em Neutrino mixing from the double tetrahedral group $T^{\prime}$,}
  arXiv:0707.3661;\\
%
  A.~H.~Chan, H.~Fritzsch and Z.~z.~Xing,
  {\em Deviations from tri-bimaximal neutrino mixing in type-II see-saw and
  leptogenesis,}
  arXiv:0704.3153;\\
%
  Z.~z.~Xing,
  {\em Nontrivial correlation between the CKM and MNS matrices,}
  Phys.\ Lett.\ B {\bf 618} (2005) 141
  [hep-ph/0503200];\\
%
  Z.~z.~Xing, H.~Zhang and S.~Zhou,
  {\em Nearly tri-bimaximal neutrino mixing and CP violation from mu - tau
  symmetry breaking,}
  Phys.\ Lett.\  B {\bf 641} (2006) 189
  [hep-ph/0607091];\\
%
  S.~K.~Kang, Z.~z.~Xing and S.~Zhou,
  {\em Possible deviation from the tri-bimaximal neutrino mixing in a see-saw
  model,}
  Phys.\ Rev.\  D {\bf 73} (2006) 013001
  [hep-ph/0511157];\\
%
  S.~Luo and Z.~z.~Xing,
  {\em Generalized tri-bimaximal neutrino mixing and its sensitivity to
    radiative   corrections,}
  Phys.\ Lett.\  B {\bf 632} (2006) 341
  [hep-ph/0509065];\\
%
  M.~Hirsch, E.~Ma, J.~C.~Romao, J.~W.~F.~Valle and A.~Villanova del Moral,
  {\em Minimal supergravity radiative effects on the tri-bimaximal neutrino
  mixing pattern,}
  Phys.\ Rev.\  D {\bf 75} (2007) 053006
  [hep-ph/0606082];\\
%
  N.~N.~Singh, M.~Rajkhowa and A.~Borah,
  {\em Deviation from tri-bimaximal mixings in two types of inverted
    hierarchical neutrino mass models,}
  hep-ph/0603189;\\
%
  X.~G.~He and A.~Zee,
  {\em Minimal modification to the tri-bimaximal neutrino mixing,}
  Phys.\ Lett.\  B {\bf 645} (2007) 427
  [hep-ph/0607163];\\
%
  N.~Haba, A.~Watanabe and K.~Yoshioka,
  {\em Twisted flavors and tri-bimaximal neutrino mixing,}
  Phys.\ Rev.\ Lett.\  {\bf 97} (2006) 041601
  [hep-ph/0603116];\\
%
  Z.~z.~Xing,
  {\em Nearly tri-bimaximal neutrino mixing and CP violation,}
  Phys.\ Lett.\  B {\bf 533} (2002) 85
  [hep-ph/0204049];\\
%
  Y.~Lin,
  {\em A predictive $A_4$ model, charged lepton hierarchy and tri-bimaximal sum
  rule,}
  Nucl.\ Phys.\  B {\bf 813} (2009) 91
  [arXiv:0804.2867];\\
%
  L.~Yin,
  {\em A dynamical approach to link low energy phases with leptogenesis,}
  arXiv:0903.0831.\\
%
  For earlier applications of discrete family symmetries see, e.g.:\\
%
  S.~Pakvasa and H.~Sugawara,
  {\em Discrete symmetry and Cabibbo angle,}
  Phys.\ Lett.\  B {\bf 73} (1978) 61;\\
%
  S.~Pakvasa and H.~Sugawara,
  {\em Mass of the t quark in SU(2) $\times$ U(1),}
  Phys.\ Lett.\  B {\bf 82} (1979) 105;\\
%
  Y.~Yamanaka, H.~Sugawara and S.~Pakvasa,
  {\em Permutation symmetries and the fermion mass matrix,}
  Phys.\ Rev.\  D {\bf 25} (1982) 1895
  [Erratum-ibid.\  D {\bf 29} (1984) 2135];\\
%
  T.~Brown, S.~Pakvasa, H.~Sugawara and Y.~Yamanaka,
  {\em Neutrino masses, mixing and oscillations in $S_4$ model of permutation
  symmetry,}
  Phys.\ Rev.\  D {\bf 30} (1984) 255.

\bibitem{Luhn:2007sy}
  C.~Luhn, S.~Nasri and P.~Ramond,
  {\em Tri-bimaximal neutrino mixing and the family symmetry $Z_7 \rtimes Z_3$,}
  Phys.\ Lett.\  B {\bf 652} (2007) 27
  [arXiv:0706.2341].

\bibitem{Lam:2008sh}
  C.~S.~Lam,
  {\em The unique horizontal symmetry of leptons,}
  Phys.\ Rev.\  D {\bf 78} (2008) 073015
  [arXiv:0809.1185].

\bibitem{King:2009mk}
  S.~F.~King and C.~Luhn,
  {\em A new family symmetry for SO(10) GUTs,}
  Nucl.\ Phys.\  B {\bf 820} (2009) 269
  [arXiv:0905.1686].

\bibitem{Grimus:2009pg}
  W.~Grimus, L.~Lavoura and P.~O.~Ludl,
  {\em Is $S_4$ the horizontal symmetry of tri-bimaximal lepton mixing?,}
  arXiv:0906.2689.

\bibitem{Lam:2009hn}
  C.~S.~Lam,
  {\em A bottom-up analysis of horizontal symmetry,}
  arXiv:0907.2206.

\bibitem{King:1998jw}
  S.~F.~King,
  {\em Atmospheric and solar neutrinos with a heavy singlet,}
  Phys.\ Lett.\ B {\bf 439} (1998) 350
  [hep-ph/9806440];\\
%
  S.~F.~King,
  {\em Atmospheric and solar neutrinos from single right-handed neutrino
    dominance  and U(1) family symmetry,}
  Nucl.\ Phys.\ B {\bf 562} (1999) 57
  [hep-ph/9904210];\\
%
  S.~F.~King,
  {\em Large mixing angle MSW and atmospheric neutrinos from single
    right-handed  neutrino dominance and U(1) family symmetry,}
  Nucl.\ Phys.\ B {\bf 576} (2000) 85
  [hep-ph/9912492];\\
%
  S.~F.~King,
  {\em Constructing the large mixing angle MNS matrix in see-saw models with
  right-handed neutrino dominance,}
  JHEP {\bf 0209} (2002) 011
  [hep-ph/0204360];\\
%
  S.~Antusch and S.~F.~King,
  {\em Sequential dominance,}
  New J.\ Phys.\  {\bf 6} (2004) 110
  [hep-ph/0405272].

\bibitem{Minkowski:1977sc}
  P.~Minkowski,
  {\em Mu $\to$ e gamma at a rate of one out of 1-billion muon decays?,}
  Phys.\ Lett.\  B {\bf 67} (1977) 421;\\
%
  M.~Gell-Mann, P.~Ramond, and R.~Slansky,
  Sanibel talk, CALT-68-709 (1979), hep-ph/9809459 (retroprint),
  and Supergravity, North-Holland, Amsterdam (1979);\\
%
  T.~Yanagida,
  Proceedings of the Workshop on Unified Theory and Baryon
  Number of the Universe, KEK, Japan (1979).

\bibitem{King:2003rf}
  S.~F.~King and G.~G.~Ross,
  {\em Fermion masses and mixing angles from SU(3) family symmetry,}
  Phys.\ Lett.\  B {\bf 520} (2001) 243
  [hep-ph/0108112];\\
%
  S.~F.~King and G.~G.~Ross,
  {\em Fermion masses and mixing angles from SU(3) family symmetry and
  unification,}
  Phys.\ Lett.\  B {\bf 574} (2003) 239
  [hep-ph/0307190].

\bibitem{Antusch:2007jd}
  S.~Antusch, L.~E.~Ibanez and T.~Macri,
  {\em Neutrino masses and mixings from string theory instantons,}
  JHEP {\bf 0709} (2007) 087
  [arXiv:0706.2132].

\bibitem{King:2009qt}
  S.~F.~King,
  {\em Tri-bimaximal Neutrino Mixing and $\theta_{13}$,}
  arXiv:0903.3199.

\bibitem{MBD}
  G.~A.~Miller, H.~F.~Blichfeldt, and L.~E.~Dickson,
  {\em Theory and application of finite groups},
  John Wiley \& Sons, New York (1916), and Dover edition (1961);\\
%
  see also P.~O.~Ludl,
  {\em Systematic analysis of finite family symmetry groups and their
    application to the lepton sector,}
  arXiv:0907.5587.

\bibitem{FFK}
  W.~M.~Fairbairn, T.~Fulton, W.~H.~Klink,
  {\em Finite and disconnected subgroups of SU(3) and their application to the
  elementary-particle spectrum},
  J.\ Math.\ Phys.\  {\bf 5} (1964) 1038;\\
%
  A.~Bovier, M.~L\"uling, and D.~Wyler,
  {\em Finite subgroups of SU(3)}
  J.\ Math.\ Phys.\  {\bf 22} (1981) 1543.

\bibitem{delta6n2}
  J.~A.~Escobar and C.~Luhn,
  {\em The flavor group $\Delta(6n^2)$,}
  J.\ Math.\ Phys.\  {\bf 50} (2009) 013524
  [arXiv:0809.0639].

\bibitem{delta3n2}
  C.~Luhn, S.~Nasri and P.~Ramond,
  {\em The flavor group $\Delta(3n^2)$,}
  J.\ Math.\ Phys.\  {\bf 48} (2007) 073501
  [hep-th/0701188].

\bibitem{Z7Z3}
  C.~Luhn, S.~Nasri and P.~Ramond,
  {\em Simple finite non-Abelian flavor groups,}
  J.\ Math.\ Phys.\  {\bf 48} (2007) 123519
  [arXiv:0709.1447];\\
%
  C.~Hagedorn, M.~A.~Schmidt and A.~Y.~Smirnov,
  {\em Lepton mixing and cancellation of the Dirac mass hierarchy in SO(10) GUTs
  with flavor symmetries $T_7$ and $\Sigma(81)$,}
  Phys.\ Rev.\  D {\bf 79} (2009) 036002
  [arXiv:0811.2955].

\bibitem{Fairbairn:1982jx}
  W.~M.~Fairbairn and T.~Fulton,
  {\em Some comments on finite subgroups of SU(3),}
  J.\ Math.\ Phys.\  {\bf 23} (1982) 1747.

\bibitem{Antusch:2008yc}
  S.~Antusch, S.~F.~King and M.~Malinsky,
  {\em Perturbative estimates of lepton mixing angles in unified models,}
  arXiv:0810.3863;\\
%
  S.~Boudjemaa and S.~F.~King,
  {\em Deviations from tri-bimaximal mixing: charged lepton corrections and
  renormalization group running,}
  arXiv:0808.2782;\\
%
  S.~Antusch, S.~F.~King and M.~Malinsky,
  {\em Third family corrections to quark and lepton mixing in SUSY models with
  non-Abelian family symmetry,}
  JHEP {\bf 0805} (2008) 066
  [arXiv:0712.3759];\\
%
  S.~Antusch, S.~F.~King and M.~Malinsky,
  {\em Third family corrections to tri-bimaximal lepton mixing and a new sum
  rule,}
  Phys.\ Lett.\  B {\bf 671} (2009) 263
  [arXiv:0711.4727];\\
%
  S.~F.~King,
  {\em Parametrizing the lepton mixing matrix in terms of deviations from
  tri-bimaximal mixing,}
  Phys.\ Lett.\  B {\bf 659} (2008) 244
  [arXiv:0710.0530];\\
%
  S.~Antusch, P.~Huber, S.~F.~King and T.~Schwetz,
  {\em Neutrino mixing sum rules and oscillation experiments,}
  JHEP {\bf 0704} (2007) 060
  [hep-ph/0702286];\\
%
  S.~Antusch and S.~F.~King,
  {\em Charged lepton corrections to neutrino mixing angles and CP phases
  revisited,}
  Phys.\ Lett.\  B {\bf 631} (2005) 42
  [hep-ph/0508044].

\end{thebibliography}
\end{document}